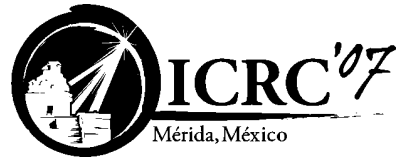

# A LED Flasher for TUNKA experiment

B.K. LUBSANDORZHIEV[1], R.V. POLESHUK[1], B.A.J. SHAIBONOV[1], Y.E. VYATCHIN[1], A.V. ZABLOTSKY[2]
[1]*Institute for Nuclear Research of Russian Academy of Science, Moscow, Russia*
[2]*Scobeltsyn Institute of Nuclear Physics of Moscow State University, Moscow, Russia*
lubsand@pcbai10.inr.ruhep.ru

**Abstract:** A LED flasher has been developed for TUNKA-133 EAS Cherenkov detector. A blue ultra bright InGaN LED is used as a light source in the flasher. The flasher's driver is based on a fast discharge of a small capacitor via a complementary pair of fast RF transistors. The light yield of the flasher is adjusted in the wide range of from 0 to up to $10^9$ photons per pulse. The results of studies of the flasher's amplitude and timing parameters and their stability are presented.

## Introduction

The history of TUNKA EAS experiment spans more than 15 years. The experiment started in the early 1990s with a toy array TUNKA-4 [1] in the picturesque Buryatian Tunka Valley in Siberia and evolved further through TUNKA-13 [2] to TUNKA-25 [3]. The basic detecting element of all TUNKA arrays has been Quasar-370G [4-6], 37 cm hemispherical hybrid phototube. Digits in the titles of the arrays indicate the number of phototubes in the arrays. TUNKA-25 has been working successfully for a number of years and is being still runned currently, giving important information on primary cosmic rays spectrum in the energy range around the classical "knee" ($\sim 3\times 10^{15}$ eV). For the time being, the development of a new array is underway at full pace.

The new array christened TUNKA-133 [7,8] will consist of 133 eight inch PMTs and cover 1 km$^2$ area. The PMTs are EMI9350 from former MACRO detector at Gran Sasso. The new array will operate in the energy range of $10^{15}$ to $10^{18}$ eV, including the classical "knee" region, measuring elaborately the primary cosmic rays energy spectrum and mass composition. The old TUNKA-25 array will continue to operate along with the new detector.

An optical detector of the new array incorporates a light source for timing and amplitude calibration of PMT. The calibration light source is based on ultra bright InGaN single quantum well (SQW) blue LED driven by a driver specially developed for this purpose.

## Ultra bright InGaN LEDs

The advent of ultra bright blue LEDs based on InGaN/GaN structures at the beginning of 1990s [9, 10] opened new era in the development of nanosecond light sources for use in different fields of experimental physics: timing and amplitude calibrations of Cherenkov and scintillator detectors, fluorescent measurements, studies of fast processes kinetics etc.

Currently a plethora of ultra bright blue LEDs are available in the market. There is a wide diversion of LEDs in their intensity and timing. Extensive studies of light yield and temporal behavior of ultra bright InGaN/GaN LEDs have been carried out by us for the last decade [11]. We have tested more than thousand of LED samples produced by more than 20 manufactures. Whereas light yields of one type of LEDs are more or less at the same level, LEDs light emission kinetics is subjected to much more variance. Even LEDs of one type produced by one manufacturer may differ very much by their light pulses profiles.

Finally we chose GNL-3014BC 3 mm single quantum well InGaN LED produced by Ningbo G-nor Opto Electronics Company. Its emission spectrum has a maximum at λ=470 nm. Albeit the LED's emission spectrum doesn't suite well to EMI9350 PMT's photocathode sensitivity curve, the LED has much higher light yield than violet LEDs. Light pulses shapes of 10 pieces of the GNL-3014BC LEDs are shown in fig.1. One can see that along with LEDs with slow emission components there are very fast LEDs without slow components tail at all.

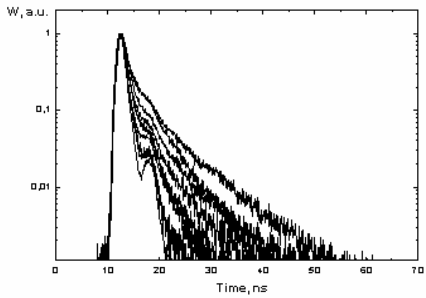

Figure 1: Light pulses shapes of GNL-3014BC LEDs produced by G-nor Electronics

We selected the fastest LEDs without slow emission components, which are capable of 1 ns width (fwhm) emission kinetics. The LED's light yield is practically the same and emission kinetics much more faster as for the famous NICHIA ultra bright blue LEDs of NSPB series. It should be noted here GNL-3014BC LEDs are extremely cheap and easily available. Among other reasons why we chose GNL-3014BC LEDs are their high reliability and good temperature behavior.

## LED driver

To drive ultra bright blue LEDs two types of electronic drivers are widely used. The highest level of light yield and at the same time the shortest light pulses are reached with a driver exploiting avalanche transistors to produce very short, a nanosecond or even less width current pulses running through LED with amplitudes of up to 3A [12, 13]. But one should be careful with electromagnetic cross talks working with such circuits. Another drawback arises from the necessity to use power supply of several hundreds of volts to feed avalanche transistors circuit and some problems with light yield adjustment.

Another type of driver is so called "Kapustinsky's driver". In 1985 J.S.Kapustinsky and his colleagues published their famous scheme of an inexpensive compact nanosecond LED pulser [14]. Since that time the pulser has become particularly popular in astroparticle physics experiments where it's widely used for time and amplitude calibrations: in high energy neutrino telescopes like NT-200 in Lake Baikal [15] and ANTARES in the Mediterranian Sea [16], the imaging atmospheric Cherenkov telescope H.E.S.S. [17], the extremely high energy cosmic detector AUGER [18] etc. The popularity resulted from the pulser's high performance, simplicity, convenience, and robustness.

We have developed our LED driver following Kapustinsky's basic scheme. The driver is based on fast discharge of a small capacitor via a complementary pair of fast RF transistors. An electrical scheme of the driver is shown in fig.2.

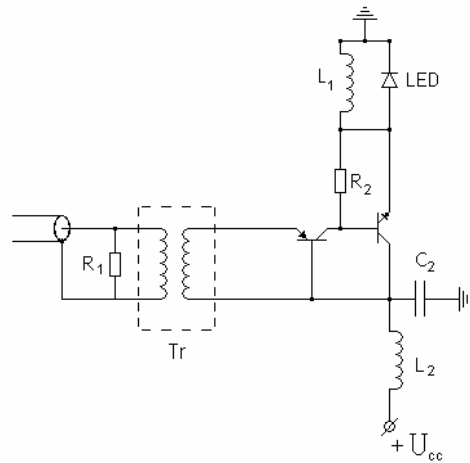

Figure 2: LED driver for TUNKA experiment

The driver is remotely operated from the cluster's electronics [19] via ~85 m coaxial cable. To avoid stray ground potential floating the driver's electronic circuit is completely isolated from the cluster's electronics by a high frequency miniature transformer. The light yield of the driver is adjusted by a simple changing of positive power supply $U_{cc}$. The driver is built on 35 mm × 35 mm PCB board. All electronic parts of the driver are easily available in the market and rather cheap. The photo of the driver is presented in fig.3.

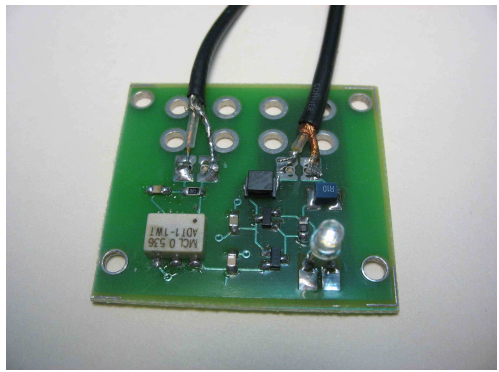

Figure 3: Photo of the driver

As it was already mentioned above, one of definite advantages of the pulser is the possiblity to adjust quite easily the light pulse intensity of LEDs by varying the power supply voltage $U_{cc}$. The latter is changed in the range of 0-12V. The light yield of the driver changes in very wide range, up to almost more than $10^9$ photons per pulse at $U_{cc}$= 12V. It's interesting that at $U_{cc}$= 12V a single light pulse from the driver is seen by just a naked eye. The light pulse width ranges from 2 ns (fwhm) at lowest value of $U_{cc}$ at which the driver light pulses start to be registered by the PMT to ~7 ns (fwhm) at $U_{cc}$= 12V.

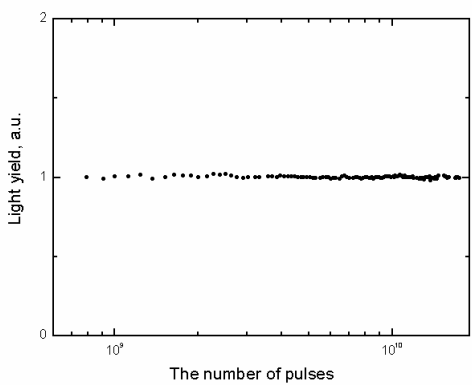

Figure 4: Long-term stability of the driver's light yield.

We measured the long-term stability of the driver's parameters. The light yield of the driver and it's light pulses width didn't practically change even after more than $10^{10}$ total pulses of driver at the highest value of $U_{cc}$ (fig.4). The light yield stability is better than 1%. The light emission kinetics of the flasher doesn't deteriorate at all in this range of the total pulses of the flasher.

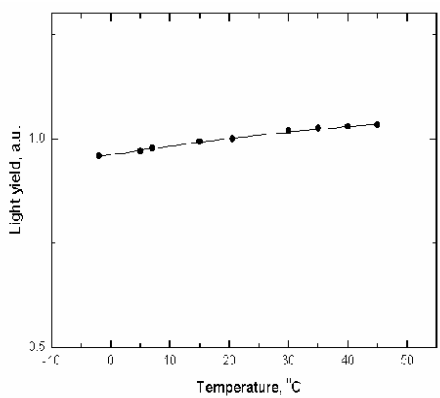

Figure 5: Temperature dependence of the driver's light yield

The typical dependence of the driver's light yield on temperature in the range of ~50°C is shown in fig.5. The measured temperature coefficient in this range is 0.15-0.2%/°C. As for the driver's light pulses width there is no any notable changes. However, in reality Tunka Valley presents temperature changes in a substantially wider range. Besides we are not going to heat inside of the optical module of the array, so it's necessary to study further the driver's parameters temperature dependence.

## Conclusion

The LED driver developed for TUNKA EAS experiment demonstrates good performances. The operation of the first cluster of the TUNKA-133 array proves the driver's reliability and robustness.

## Acknowledgements

The authors are very much indebted to Prof. L.B.Bezrukov for help at every stages of the driver's development.